\documentclass[aps,preprint,tightenlines]{revtex4}
\usepackage{amssymb,amsmath}
\def\beq{\begin{equation}}
\def\eeq{\end{equation}}
\def\q{\mathbf{q}}
\def\p{\mathbf{p}}
\begin{document}
\title{Constraints and gauge transformations: Dirac's theorem is not always valid}
\author{Julian Barbour}
\email[]{julian.barbour@physics.ox.ac.uk} \affiliation{College Farm,
The Town, South Newington, Banbury, OX15 4JG, UK}
\author{Brendan Z. Foster}
\email[]{B.Z.Foster@phys.uu.nl} \affiliation{Institute for
Theoretical Physics, Utrecht University, Leuvenlaan 4, 3584 CE
Utrecht, The Netherlands}
\preprint{ITP-UU-08/46, SPIN-08/36}

\date{August 8, 2008}
%
%
\begin{abstract}
A standard tenet of canonical quantum gravity is that evolution
generated by a Hamiltonian constraint is just a gauge
transformation on the phase space and therefore does not change
the physical state. The basis for this belief is a theorem of
Dirac that identifies primary first-class  constraints as
generators of physically irrelevant motions.  We point out that
certain assumptions on which Dirac based his argument do not hold
for reparametrization invariant systems, and show that the primary
Hamiltonian constraint of these systems does generate physical
motion. We show explicitly how the argument fails for systems
described by Jacobi's principle, which has a structure closely
resembling that of general relativity.  We defer discussion of
general relativity and the implications for quantum gravity to a
later paper.
\end{abstract}
\maketitle
%
%
\section{Introduction}
A standard belief in the canonical quantum gravity community is
that the evolution generated by any Hamiltonian constraint $H$
``is just the unfolding of a gauge
transformation"~\cite{Henneaux:1992ig}. This suggests that if
observables in quantum gravity are gauge invariant, they must
commute with $H$---they must be `perennials', to use Kucha\v r's
coining~\cite{Kuchar:1992gs}. Kucha\v r argues persuasively,
however, that this feature runs counter to intuition; he therefore
makes the valuable distinction between what he calls
\emph{observables}, which need only commute with the ADM momentum
constraint, and the perennials, which commute with both  it and
$H$. He also notes that perennials are extraordinarily difficult
to find in any dynamical system, let alone one as intricate as
general relativity (GR). Kucha\v r~\cite{Kuchar:1992gs} and one of
us~\cite{Barbour:1994ri} have also argued that the physical state
is manifestly changed by constrained Hamiltonian evolution. In
this paper, we will sharpen the challenge to the standard belief
by showing that the premise on which it is based is false.

The standard designation of the GR Hamiltonian as a gauge generator
is based on a simple argument of Dirac~\cite{Dirac}, from which he
concluded that primary first-class constraints treated as generating
functions of infinitesimal contact transformations ``lead to changes
in the [canonical data] that do not affect the physical state''.
Thus, one concludes that all points in the phase space on an orbit
of such a constraint represent the same physical state but in
different gauges; the constraint implements a gauge transformation
in carrying the state from one point to another along such an orbit.
This interpretation is universally accepted for, say, the Gauss
constraint in electrodynamics. $H$ is a primary first-class
constraint in GR, so it has been widely concluded that it too is a
gauge generator of this kind.

The symmetry connected to $H$ in GR is invariance of the action
under a change of foliation of a single spacetime, together with
the freedom to change the curve parameter that labels the leaves
of a fixed foliation. It is not widely recognized that a
refoliation and a reparametrization are distinct operations, but
the difference is indicated by the fact that the former changes
the curve in phase space, while the latter does not. A theory is
reparametrization invariant (RI) if and only if its Lagrangian is
homogeneous of degree one in its velocities. Such a theory
necessarily has a primary first-class  constraint: its Hamiltonian
$H$ must vanish.

In this paper, we shall not consider fully foliation invariant
theories, but just those that are simply RI and, for simplicity,
do not have any other constraints apart from $H=0$.  We will not
include theories with `$1\times\infty$' constraints---that is,
field theories with a single local constraint. Parametrized
particle dynamics is often invoked as a simple RI analogue of GR.
Jacobi's principle in classical dynamics, however, leads to an RI
theory that is a much better
analogue~\cite{Barbour:1994ri,Barbour:2000qg}. We shall use it
here to show that Dirac's argument \textit{does not have general
validity.} Our argument will apply to all theories of  the class
that we consider, including parametrized dynamics.

The essential flaw in Dirac's argument is this: His proof, given
in Chp.~1 of his \emph{Lectures on Quantum
Mechanics}~\cite{Dirac}, assumes that the independent variable of
the Hamiltonian evolution is an absolute time. In Chp.~3, Dirac
then considers RI theories and shows that in them there is no
absolute time, but he does not consider the implications for his
earlier proof. In particular, he does not repeat the explicit
calculation that is the key to his result. We shall show, by
properly completing this calculation, that the difference
invalidates the proof for RI theories. In their standard reference
on constrained systems, Henneaux and
Teitelboim~\cite{Henneaux:1992ig} do note that the issue of the
status of time affects the argument but still adopt the standard
conclusion. We will review the discussions of~\cite{Dirac}
and~\cite{Henneaux:1992ig} below.

We emphasize that our claim is much more significant than a
refutation of the `Dirac conjecture' that all \emph{secondary}
first-class constraints generate motions that do not change the
physical state. Dirac makes this conjecture in Chp.~1~\cite{Dirac}
just after giving his proof for \emph{primary} first-class
constraints. The Hamiltonian constraint in GR and RI theories is a
primary constraint, so it is Dirac's apparent proof that we are
concerned with. See~\cite{Henneaux:1992ig} for more discussion of
the conjecture, including a simple counterexample.

Nearly all quantum gravity reviews invoke the standard tenet with
no discussion of the details. The only serious challenge to the
`orthodoxy' that we have seen is from
Kucha\v{r}~\cite{Kuchar:1992gs}, although
neither~\cite{Kuchar:1992gs} nor~\cite{Barbour:1994ri} makes clear
how the problem is resolved by identification of the weak points
in Dirac's argument. In his recent book,
Kiefer~\cite{Kiefer:2004gr} does present Dirac's argument, though
without identifying the assumption of absolute time; he therefore
concurs with the standard conclusion but emphasizes the importance
of Kucha\v r's distinction between observables and perennials,
thus avoiding the counterintuitive implication of the standard
conclusion.

Our result casts strong doubt on the standard quantum gravity
belief. We speculate here that the Hamiltonian constraint in GR is
partly a gauge generator and partly a generator of true physical
change---and moreover more the latter than the former. Because full
foliation invariance is a much more complex issue than simple
reparametrization invariance, we defer further discussion of it to a
later paper. We will also consider there the difficult question of
whether observables in quantum GR must be perennials. We are however
confident that they need not be for simple RI theories, as we shall
also argue in a later paper.

We should also emphasize that we follow closely Dirac's method of
investigation and consider the infinitesimal change of the state
generated by an infinitesimal action of the Hamiltonian constraint
in the case of RI theories. So far as we know, no one has done this
hitherto when considering the meaning of constraints in RI theories;
Dirac certainly did not and therefore failed to follow the logic of
his own method. As a consequence, his result, which is correct for a
large class of theories with other kinds of constraints, has been
widely applied without question to RI theories.

In this connection, some argue---incorrectly in our view---that
the entire history in phase space, rather than the individual
points on it, defines the physical state~\cite{Rovelli:2004tv}. It
has also been suggested to us in correspondence that the
Hamiltonian constraint could be considered to act on this complete
history and thereby generate a reparametrization of it. But
reparametrization of an entire history is quite different from the
infinitesimal action of the constraint on the instantaneous state,
which is what we need to consider in following the logic of
Dirac's approach. We believe that faithful adherence to that logic
will at last bring clarity to this vexed issue.

We will now go into more detail, beginning with a discussion of the
nature of time itself and its relation to Jacobi's principle and
parametrized dynamics. We then present Dirac's argument, identify
its fault, and summarize what was actually said by Dirac, and
Henneaux and Teitelboim.  It is convenient to first study Jacobi's
principle, as it makes the arbitrariness of the curve parameter and
the definiteness of the Hamiltonian evolution clear. This
preparation demonstrates beyond doubt that Dirac's result does not
have general validity and that the Hamiltonian constraint in this
class of theories generates physical motion. Throughout, we follow
Dirac's conventions~\cite{Dirac}.
%
\section{Time and Jacobi's principle}\label{SEC:TWO}
%
\subsection{Time}\label{SEC:TIME}

We begin with comments about time itself. Newtonian dynamics treats
it as absolute: an external independent variable unrelated to what
happens in the universe. In the Newtonian $n$-body problem, a
dynamical history is a curve in a $3n$-dimensional configuration
space Q traversed at a speed measured by that time $t$. But we could
never say that time passes if we did not observe change in general
and motion in particular. In reality, time is measured by a motion;
for millennia the rotation of the earth provided the hand of the
clock and the fixed stars the clock face. In a universe consisting
of $n$ particles with no nearly rigid earth and fixed stars, all
that is given objectively is the curve in Q. Deprived of an easy
convenient measure of time, we must think about these things
clearly. What is time? How do we express the objective content of
Newtonian theory without absolute time?

In fact, we shall show that the curve in Q contains all that is
essential: the succession of configurations that the Newtonian $t$
labels. What is lost is the metric of $t$, which becomes a mere
label better called $\lambda$, and the direction of time. But
neither are essential. We shall see shortly that the true content of
Newtonian dynamics is that it determines such curves given an
initial point $q$ in Q and an initial direction at $q$, which we
denote $d_q$. The physical state is defined by the pair $(q, d_q)$.

Before we present the theory of geodesics on Q, let us note that it
will meet not only the epistemological insight that time must be
derived from change but also the requirement that a dynamical curve
be determined by a pair $(q, d_q)$. Such a pair constitutes the
initial condition for a geodesic. Note also that specification of
$q$ requires $3n$ numbers but of $d_q$ one less. We shall see that
this is the natural and necessary origin of Hamiltonian constraints.

Because the geodesic theory we shall obtain is a representation of
Newtonian theory defined on the same space Q, distinct points in Q
must represent physically distinct states just as they do in
Newtonian theory. Indeed, we can assert that the same is true for
points in the configuration space for all RI theories with just
the single Hamiltonian constraint. Thus all the considered
dynamical variables are observable. As we explain below, this
includes the time in parametrized dynamics; only the unphysical
label need be assumed to be unobservable.

We could now proceed to construct a timeless theory of geodesics in
Q \emph{ab initio} and show how Newtonian theory can be recovered
from it. It will be more useful for the purposes of our discussion
to work in the opposite direction, however, which we do now.
%
\subsection{Jacobi's principle}\label{SEC:JACOBI}
%
Consider an action $I$, evaluated along a fixed curve in Q with
end point configurations  $(A,B)\in$ Q:
\beq\label{ACTION}
    I = \int^B_A d\lambda L[q_i,\dot q_i],
\eeq
where $\dot q_i = dq_i/d\lambda$.  Consider the case where $L$ is
homogeneous of degree one in the velocities $\dot q_i$. Under a
reparametrization of the curve label $\lambda$,
\beq\label{REPR}
    \lambda \rightarrow\lambda'(\lambda),
\eeq
$L$ transforms as
\beq
    L[q_i,\frac{dq_i}{d\lambda}]\rightarrow L'[q_i,\frac{dq_i}{d\lambda'}] =
            \frac{d\lambda}{d\lambda'} L[q_i,\frac{dq_i}{d\lambda}].
\eeq
Thus, the action~\eqref{ACTION} is invariant under the
reparametrization~\eqref{REPR}; $\lambda$ is a mere label without
physical significance.  Note that it is not necessary to leave
fixed the value of $\lambda$ at the end points of integration. To
express the invariance of the action properly requires stipulating
fixed configurations $A$ and $B$; it is meaningless to fix the end
points by giving values of $\lambda$.

The general form of $L$ also implies  that the canonical momenta
$p_i = \delta L/\delta \dot{q}_i$ are homogeneous of degree zero and
thus invariant under a reparametrization.  Furthermore, the
canonical Hamiltonian vanishes identically:
\beq\label{HAMIDENT}
    \sum_i p_i \dot{q}_i - L = 0.
\eeq
This `Hamiltonian constraint', expressed as a function of the
canonical data $(q_i,p_i)$, is a primary first-class constraint;
it is thus the type to which Dirac's argument will apply. A
\emph{primary} constraint is a relation between the canonical
variables that holds by definition of the momenta.
Obviously,~\eqref{HAMIDENT} is of this type. A \emph{first-class}
function of the canonical variables is one such that its Poisson
bracket with every constraint of the system is a linear
combination of the constraints.  Since we will only consider
systems with just this Hamiltonian constraint, the constraint is
first-class.

Let us consider the explicit example of a system of Newtonian
particles described by Cartesian coordinates $\q_i$. We can begin
with a standard, non-RI action $I$, defined along a curve in Q
with fixed end point configurations $A$ at time $t_0$ and $B$ at
$t_f$:
\beq\label{NEWTI}
    I = \int_{A(t_0)}^{B(t_f)} dt
        (\frac{1}{2}\sum_i \frac{d\q_i}{dt}\cdot\frac{d\q_i}{dt}-V(\q_i)),
\eeq
where all the particle masses are set to unity for simplicity. We
render this action RI by promoting the time coordinate $t$ to a
dynamical variable and introducing an evolution parameter
$\lambda$ that labels the curve in the new configuration space
$\textrm{Q}T$, Q augmented by the space $T$ of absolute times. The
action becomes
\beq
    I = \int_{(A,t_0)}^{(B,t_f)} d\lambda (\frac{T}{\dot{t}} -
    \dot{t}V),\label{PARAMETRIZED}
\eeq
where $T=\frac{1}{2}\sum_i \dot{\q}_i\cdot\dot{\q}_i$. The canonical
momentum $p_t$ of $t$ is
\beq
    p_t = -(\frac{T}{\dot{t}^2}+V),
\eeq
while the canonical momentum $\p_i$ of particle $i$ is
\beq
    \p_i = \frac{\dot{\q}_i}{\dot t}.
\eeq
The identity~\eqref{HAMIDENT} can be expressed as
\beq\label{HAMIDENT2}
    \frac{1}{2}\sum_i \p_i \cdot\p_i + p_t + V = 0.
\eeq

Since $T/(\dot t)^2$ is the unparametrized kinetic energy, $p_t$
is just minus the total energy $E$.  Because $t$ is cyclic ($L$
depends only on its velocity $\dot t$), $p_t$ is conserved.  We
can then eliminate $\dot t$ from the action via Routhian
reduction. Following standard procedure~\cite{Lanczos}, we do this
by solving $p_t = -E$ for $\dot t$ and substituting the result
into a reduced Lagrangian $\bar L$:
\beq
    \bar L = L + E\dot{t}.
\eeq
This leads to the reduced, `Jacobi' action $I_J$:
\beq\label{JACOBI}
    I_J = 2 \int_A^B d\lambda \sqrt{(E-V) T},
\eeq
with a characteristic square root form. $I_J$ defines a geodesic
principle, Jacobi's principle, on the original configuration space
Q.

The timeless nature of Jacobi's principle is especially clear when
expressed without an evolution parameter, using only the $\q_i$ and
their displacements $\delta \q_i$:
\beq\label{NOTIME}
    I_J = \int_A^B \sqrt{2(E-V)\sum_i\delta \q_i\cdot\delta \q_i}.
\eeq
This is the timeless geodesic theory that we could have derived
\emph{ab initio} based on the inescapable fact that time as such
is not observable. Here, $E$ is regarded as part of the
potential---that is, part of the law of the `island universe'
described by (\ref{NOTIME}); Einstein's cosmological constant
$\Lambda$ occurs in the Baierlein--Sharp--Wheeler action for
general relativity in exactly the same way~\cite{Barbour:2000qg}.
The form~\eqref{NOTIME} of the action makes the physical
irrelevance of any label $\lambda$ evident and shows that the
curve end points for the calculation of the action are defined by
the configurations $A$ and $B$.
%
\subsection{The Lapse}\label{SEC:LAPSE}
%
In conventional applications, Jacobi's principle is used to split
dynamical problems into the determination of a timeless orbit in the
configuration space; energy conservation is then used to determine
the speed in orbit. However, if the system under consideration
models the universe, it is more illuminating to see how a physical
and directly observable time along the orbit can be
obtained~\cite{Barbour:1994ri}. Define the `lapse' function $N$:
\beq\label{DEFN}
    N = \sqrt{\frac{T}{E-V}}.
\eeq
Then the canonical momenta $\p_i$ corresponding to Jacobi's
principle (\ref{JACOBI}) are
\beq\label{DEFP}
    \p_i = \frac{\dot{\q}_i}{N},
\eeq
and the Euler--Lagrange equations are
\beq
    \dot{\p}_i = -N \frac{\partial V}{\partial \q_i}.
\eeq
The canonical Hamiltonian $H$ vanishes identically, but we can
formally express it:
\beq\label{JACOBIH}
    H = \sum_i \p_i\cdot \dot{\q}_i - \bar{L}
        = N h,
\eeq
where
\beq
    h\equiv\frac{1}{2} \sum_i \p_i\cdot \p_i + V -E,
\eeq
vanishes by the definitions~\eqref{DEFN} and~\eqref{DEFP}. The
Hamiltonian constraint $h=0$ arises because the initial condition
for a geodesic principle involves a direction and not a direction
with a speed along it. Directions are most conveniently specified by
direction cosines, which satisfy a constraint: the sum of their
squares is unity. The Jacobi momenta are essentially direction
cosines multiplied by $\sqrt{(E-V)}$, so their squares sum to give
the quadratic identity.

The definitions~\eqref{DEFN} and~\eqref{DEFP} also show that a
given set of canonical data $(\q_i,\p_i)$ such that the $\p_i$
satisfy the Hamiltonian constraint are \emph{uniquely determined}
by the physical state $(q,d_q)$, and vice versa. Thus, $H$
generates a unique curve of the evolution in phase space, and each
point along it corresponds to a unique physical state. The only
`gauge' aspect of this description is the value of the unphysical
label at which a particular state is reached.

A crucial property of the lapse is that it transforms $N\rightarrow
N'$ under a reparametrization such that
\beq\label{NTRANS}
     N'(\lambda')d\lambda' =  N(\lambda)d\lambda.
\eeq
This result is obvious from the definition~\eqref{DEFN}, but it can
also be derived in the Hamiltonian picture---that is, without
invoking the definitions of lapse~\eqref{DEFN} or
momenta~\eqref{DEFP}---by requiring invariance of the canonical
action $I = \int d\lambda (\sum_i \p_i \dot\q_i - H[\q,\p])$.  The
transformation of $I$ must again be made with fixed curve end points
for the invariance to be physically meaningful. The lapse is an
arbitrary function, in the sense that its value is not fixed by the
dynamical equations; however, given a choice of lapse and
parametrization of a dynamical curve, changing the lapse implies
changing the parameter in the manner of~\eqref{NTRANS}.

The lapse also enables us to recover Newtonian time and to see how
it arises from the genuinely observable $\q_i$ and $\delta\q_i$.
Requiring that the Euler--Lagrange equations take the form of
Newton's laws leads to the condition $N=1$, equivalent to a choice
of distinguished parameter $t$. The increment $\delta t$ of this
Newtonian time corresponds to
\beq\label{NEWTIME}
    \delta t = N \delta\lambda
        =\sqrt{\frac{\sum_i \delta \q_i\cdot \delta \q_i}{2(E-V)}}.
\eeq
Thus, the increment of physical time $\delta t$ is determined by the
shape of the geodesic.

We can also now see the conceptual relationship between Jacobi's
principle and parametrized dynamics. Suppose a system of $n$ point
particles and an effectively isolated subsystem of $m$ particles
within it ($n\gg m$). The time $t$ with increment $\delta
t=Nd\lambda$ can be obtained for the large system as we have done
for Jacobi theory and then serve as the independent variable for
the small subsystem, which can then be parametrized. This makes it
clear that the time $t$ in parametrized dynamics is in principle
as observable as the $q$'s and $p$'s of the subsystem to which it
is adjoined for the purposes of parametrization. It may be worth
mentioning in this connection, however, that the time obtained in
this manner is not a function on the phase space. This is because
the many-to-one relationship between the velocities and the
momenta expressed by the Hamiltonian constraint precludes
conversion of the $\delta{\textbf{q}_i}$ in (\ref{NEWTIME}) into
momentum-type variables.

Finally, we note that Dirac's methods~\cite{Dirac} can be used to
show that the Hamiltonian for the general RI theories that we
consider has the form~\eqref{JACOBIH} as in Jacobi theory. $N$ will
again be an arbitrary function, and invariance of the canonical
action will imply the invariance of $N d\lambda$. The function
$h(p,q)$ will be a primary first-class constraint, although
potentially with a different form.  In parametrized particle
dynamics, for example, $h$ is given by~\eqref{HAMIDENT2}.

%
\subsection{Evolution}\label{SEC:EVOL}
%
Evolution in the case of RI theories is quite unlike ordinary
unconstrained evolution.  We will focus on evolution associated
with an infinitesimal amount of action, because this is the case
Dirac considers.  In standard Hamiltonian theory in which absolute
time $t$ is the integration variable, one can generate a physical,
infinitesimal motion by letting $H$ act for infinitesimal $\delta
t$. In RI theories, however, an infinitesimal $\delta\lambda$ has
no meaning by itself. In fact, we have seen in~\eqref{NOTIME} that
Jacobi theory can be defined with no $\lambda$ whatsoever. The
only meaningful increment is of the emergent Newtonian time
$\delta t$~\eqref{NEWTIME}.  If we use a parameter, the increment
of evolution only becomes well-defined when we specify both
$\delta\lambda$ and $N(\lambda)$. It should also be noted that a
finite evolution is obtained by specifying $N(\lambda)$ over some
finite interval $[\lambda_1,\lambda_2]$ of the independent
variable. The action of $H$ then generates the curve in phase
space \emph{and} the parametrization on it. Therefore, one cannot
say that $H$ generates a \emph{re}parametrization since no
parametrization exists before $H$ acts.

The infinitesimal change of any dynamical variable $g$ along the
evolution curve is
\beq\label{HAMEVOL}
    \delta g \approx \delta\lambda [g,H] = \delta t[g,h],
\eeq
where the brackets denote the Poisson bracket  and $\approx$ denotes
Dirac's weak equality~\cite{Dirac}, meaning that the relation holds
provided the Poisson bracket is calculated before $h=0$ is imposed.
For a chosen $N$ and $\lambda$, we see that $\delta g$ is
proportional to $N\delta\lambda = \delta t$, so that it is a
meaningful difference. In particular, $\delta g$ is generally
nonvanishing and is unchanged under a reparametrization.

We see that although the evolution is driven by a Hamiltonian
constraint, there is no doubt that the evolution carries us to
different configurations $q\in$Q. As remarked in
Sec.~\ref{SEC:TIME}, these  configurations must represent physically
distinct states.
Of course, if we ask what state is reached at a particular value of
the curve label $\lambda$, the answer will be changed by
reparametrization. But it is the product $N\delta\lambda$ which
determines an infinitesimal Hamiltonian evolution. This, and with it
the physical evolution, is invariant under a reparametrization that
respects the symmetry of the canonical action.
%
%
\section{The standard argument}
We have just argued that, counter to the standard belief, the
Hamiltonian constraint in single-constraint RI theories generates
physical motion. We will now present Dirac's `proof' that primary
first-class constraints generate transformations that do not change
the physical state, but demonstrate precisely how it fails for these
theories.
\subsection{Dirac's first lecture}
We begin by noting that nowhere in his \emph{Lectures} does Dirac
use the terms `gauge transformation' or `gauge generator',
presumably because gauge theory was not the dominant paradigm in
theoretical physics at the time.  Instead, he considers various
kinds of constraints and their effects on a dynamical system.
Thus, for Dirac, the precise question is not whether a constraint
is a `gauge generator' but whether it generates motions in the
phase space ``that do not affect the physical state."

In Chp.~1 of his \emph{Lectures}~\cite{Dirac}, Dirac considers
systems with a `true' Hamiltonian $H$, which is not constrained to
vanish, but when some constraints are present. He shows that then
the system is governed by a total Hamiltonian $H_T$ of the form
\beq
    H_T=H+v_a\phi_a,
\eeq
where $\phi_a$ are primary first-class constraints (as defined
below our Eqn.~\eqref{HAMIDENT}) and $v_a$ are arbitrary functions
of the time $t$. (We have here ignored the difference in Dirac's
Eq. (1-33) between $H$ and his $H'$, which is unimportant in this
context.) The increment in a dynamical variable $g$ in the time
$\delta t$ is [Dirac's (1-37)]:
\beq\label{FULLEVOL}
    \delta g = \delta t[g,H_T] = \delta t\{[g,H]+v_a[g,\phi_a]\}.
\eeq
Dirac shows that the arbitrariness of $v_a$ allows one to change
$\delta g$ freely via a new choice $v'_a$, with a difference of:
\beq\label{AMBIG}
    \Delta g(\delta t)=\varepsilon_a[g,\phi_a],
\eeq
where
\beq
     \varepsilon_a=\delta t(v_a-v_a'),
\eeq
``is a small arbitrary number, small because of the coefficient
$\delta t$ and arbitrary because the $v$'s and $v'$'s are
arbitrary.'' Dirac assumes that the theory is such that ``the
initial state must determine the state at later times"; thus, the
ambiguity~\eqref{AMBIG} in the evolution~\eqref{FULLEVOL} implies
that the $\phi_a$ are ``\emph{generating functions [that] lead to
changes in the q's and the p's that do not affect the physical
state"} (Dirac's italics).

Dirac's ability to draw this conclusion depends crucially on the
presence of the two terms of very different nature present in
(\ref{FULLEVOL}). They are present simultaneously because there is
not only the `true' Hamiltonian $H$, which unquestionably
generates physical motion,  but also the constraints. The presence
of the absolute time $t$ is also vital: Dirac assumes that
evolution with different $v_a$ but over an identical increment
$\delta t$ gives the same physical state. Subject to these
important qualifications, we can see no flaw in Dirac's argument
as presented. Henneaux and Teitelboim~\cite{Henneaux:1992ig} give
essentially the same argument.
\subsection{Dirac's third lecture}
In his third lecture, Dirac considers how his theory of generalized
Hamiltonian dynamics can be modified to take into account
relativity. For this, he says that we cannot simply use the ``one
absolute time" assumed in the first lecture.  We must now ``go back
to first principles. It is no longer sufficient to consider just one
time variable referring to one particular observer\ldots We want to
have the possibility of various times which are all on the same
footing."

To indicate how this can be done, he considers the special case in
which the Lagrangian is homogeneous of degree one in the velocities.
The total Hamiltonian is now built up entirely from primary
first-class constraints with arbitrary coefficients:
\beq\label{HAMT}
     H_T=v_a\phi_a,
\eeq
``showing that there must be at least one primary first-class
constraint if we are to have any motion at all''. We shall attempt
to interpret the quoted words in a moment.  Note that while Dirac
wishes ultimately to consider theories that are relativistic in the
usual sense, his general analysis includes the nonrelativistic
theories that we have discussed above.

Dirac notes that the equations of motion are now
\beq\label{MOTION}
    \frac{dg}{d\lambda}\approx v_a[g,\phi_a].
\eeq
He notes that multiplying the arbitrary functions $v_a$ by a common
factor is equivalent to introducing another time variable $\lambda'$
and gives the new equations of motion
\beq\label{NEWMOTION}
    \frac{dg}{d\lambda'}\approx v_a'[g,\phi_a],
\eeq
noting that now ``there is no absolute time variable."

Now comes the crux. Having made this correct observation, Dirac then
moves on to another issue without doing the calculation analogous to
(\ref{FULLEVOL}) for the increment $\delta g$ in the case of
theories with homogeneous velocities. Had he done so, he would have
obtained the expression~\eqref{HAMEVOL}
\beq\label{NEWEVOL}\begin{split}
    \delta g &= \delta\lambda(dg/d\lambda)
            \\&=\delta t [g,h],
\end{split}\eeq
where we have specialized to $H$ of single-constraint RI
theories~\eqref{JACOBIH} in the second line.

Whereas the expression for evolution in the first
lecture~\eqref{FULLEVOL} contains two terms, the first corresponding
to physical change with an unknown admixture of unphysical change,
and the second to pure unphysical change,~\eqref{NEWEVOL} contains
only one term. This expression must represent a pure physical change
to the state, since the right-hand side is nonvanishing for general
$g$, and distinct sets of canonical data in Jacobi theory (as in all
theories of the class we consider) represent physically distinct
states, as we have noted in Sec.~\ref{SEC:LAPSE}.

Of course, as we noted earlier, one can get different values for
$\delta g$ by changing $N$ while fixing $\delta\lambda$, but one
always gets physical change. Moreover, it is clearly confusing the
issue to not change $\delta\lambda$ and $N$ simultaneously.
Transforming them both so as to respect the symmetries of the action
means leaving $\delta t$ invariant, and thus also $\delta g$.
\subsection{Dirac's commentary}
To be clear, Dirac does not explicitly state anywhere in his
\emph{Lectures} that the Hamiltonian constraint in the theories he
considers generates only physically irrelevant motions, despite
frequent mentions that it is a primary first-class constraint. In
the light of the emphasis placed in the first lecture on his proof,
however, we find it remarkable that he nowhere discusses the
apparent discrepancy between that result and the implication
that~\eqref{HAMT} does generate real change even if the `speed'
depends on an arbitrary label. What else can Dirac's words ``if we
are to have any motion at all'' mean?

Several passages in Dirac's third lecture suggest that he
instinctively took~\eqref{MOTION} and other analogous equations to
represent real change. Perhaps the most indicative comment comes
in the midst of his discussion of the Hamiltonian structure of a
parametrized field theory on flat spacetime.  Dirac has extended
the parametrization to include the four components of the
Minkowskian coordinate system. This allows him to describe
evolution along a sequence of curved surfaces, rather than just
the customary flat ones.  As a result, the Hamiltonian is just the
sum of four local constraints. Dirac demonstrates that three
``tangential" constraints generate coordinate transformations
within a given surface while the fourth ``normal" constraint
generates deformations of the surface normal to itself.

He then states that the tangential components ``[are] not of real
physical importance."  By contrast, ``the quantity which is of real
physical importance is the normal component." The normal motion that
it generates ``is something which is of dynamical importance."  For
someone who typically chose his words very carefully, ``physically
important" must surely be distinct from ``not physically important".

We have also examined Dirac's earlier paper~\cite{Dirac:1950pj} in
which he first developed his generalized Hamiltonian dynamics.
This paper puts even more emphasis than~\cite{Dirac}  on
Lagrangians that are homogeneous of degree one in the velocities,
``because they are specially convenient for a relativistic
treatment."  First- and second-class constraints are defined
explicitly, but the distinction between primary and secondary
constraints is not highlighted and Bergmann's terminology for them
is not introduced. There is nothing analogous to the proof in the
\emph{Lectures} that primary first-class constraints generate
transformations of the canonical data that do not change the
physical state.

We can only conclude that, in a remarkable oversight and because
his main interest in Chp. 3 lay in the applications of RI
theories, Dirac failed to see the contradiction between the action
of Hamiltonian constraints and his theorem. He may also have been
misled by the situation that arises in relativistic
(foliation-invariant) systems. Here arbitrary elements certainly
do come into play, and it is not easy to separate the genuine
change from irrelevant change. This is why we considered it
important to begin with the presentation of non-relativistic
Jacobi theory, for which the case is much clearer.
\subsection{Henneaux and Teitelboim's commentary}
The book~\cite{Henneaux:1992ig} of Henneaux and Teitelboim (HT) is
perhaps the best known and most thorough modern discussion of
Dirac's methods for constrained systems.  It is thus an influential
text, and we should examine how it has helped to strengthen the
widely held view that Hamiltonian constraints are to be regarded as
gauge generators. We shall see that HT state conditions that
indicate a possible invalidity of Dirac's argument for Hamiltonian
constraints but still argue for its general validity.

HT present Dirac's argument in their Chp.~1, at the end of which
they add a word of caution, commenting that the argument implicitly
assumes ``that the time $t$ ... is observable.'' They say that this
is ``information brought in from outside'' and suggest one ``may
also take the point of view that some of the gauge arbitrariness
indicates that the time itself is not observable''. The whole of
their Chp.~4 is then devoted to this issue.

There, we note that HT, like Dirac, make no attempt to pose or
answer the question that ought to be addressed first of all: What is
time? There is no suggestion that change is primary and time as a
measurable quantity can emerge through a Jacobi-like timeless law
that describes change. Instead, they comment that normally ``time is
assumed to have a direct physical significance but is not itself a
dynamical variable''. They then mention ``a different formulation
... in which the physical time and the dynamical variables ... are
treated more symmetrically''. This is just parametrized dynamics,
which they present later in the chapter. They then note that
parametrized dynamics can be readily deparametrized, but not GR.

Then comes the decisive passage, whose conclusion we question. HT
say that because attempts to deparametrize GR ``have not been quite
successful'', it seems [our italics]
\begin{quote}
preferable to aim at both formulating and answering questions while
treating all variables on the same footing. This coincides with the
point of view about gauge invariance that has been taken throughout
this book. It is not accidental that this should be so, since, as we
shall see below, \emph{in a generally covariant theory the motion is
just the unfolding of a gauge transformation}.
\end{quote}
The `proof' of this statement appears to consist of three lines in
the discussion of parametrized dynamics, when they note that its
Hamiltonian consists solely of a constraint and ``therefore, the
motion is the unfolding of a gauge transformation''.

However, this is not a proof---it is an assertion. The issue is not
whether the Hamiltonian is a constraint, which of course it is, but
\emph{what the constraint does}. As we have seen, it generates real
change in all theories of the class we consider, including
parametrized dynamics. HT's conclusion would only follow if they had
already proved that all primary first-class constraints generate
gauge transformations. Thus, their Chp.~4 purports to extend the
correct proof when time is absolute to all cases, but when it comes
to the critical point the result is simply asserted without
foundation.

We have interpreted HT's words, ``the motion is the unfolding of a
gauge transformation," to mean that $H$ carries any state on which
it acts into other states that are physically identical. This is
by analogy with the familiar Gauss constraint and in accordance
with the framework of Dirac's argument. However, immediately after
the above assertion, HT discuss an infinitesimal reparametrization
of a finite segment of the dynamical orbit in a manner suggestive
of the alternative interpretation that $H$ is supposed to perform
a transformation of an existing \emph{parametrization} of part or
all of the system history. But as we pointed out at the start of
Sec. IID, $H$ generates the physical evolution and simultaneously
some parametrization for it. It cannot therefore implement a
\emph{re}parametrization.  In any case, as we have already noted,
the label $\lambda$ has no physical significance whatsoever. All
the difficulties surrounding the status of $H$, in RI theories at
least, are immediately resolved by the simple observation that
Dirac's theorem does not hold for them.

We also comment that ``the motion is the unfolding of a gauge
transformation'' is a categoric statement, the like of which is
not found in Dirac's \emph{Lectures}. As we have noted, for some
reason or other, Dirac never uses such language or anything
equivalent to it when discussing Hamiltonian constraints even
though, as primary first-class constraints, one might have
expected him to apply his `proof' to them.
%
\section{Conclusion}
%
It is our conviction that much confusion has entered the
discussion of the meaning of constraints because of the manner in
which Dirac chose to develop his theory of them.  The great
strength of his theory is its wonderful generality: the
mathematical formalism applies with complete rigor to any
Lagrangian with a many-to-one relationship between the velocities
and the canonical momenta.  This leads inevitably to the
appearance of constraints in the Hamiltonian representation of the
theory.  The mathematical results that follow from this are not in
doubt, and their utility has been proved again and again,
especially in connection with the property of constraints being
first or second class, which has come to be seen as the
distinction \emph{par excellence} between constraints.

It is however striking that Dirac, followed by HT, makes virtually
no attempt to explain \emph{why the constraints occur in the first
place}. Speaking of the arbitrary functions that appear in the
solutions of the equations of motion due to the presence of the
constraints, Dirac says: ``These arbitrary functions of the time
must mean that we are using a mathematical framework containing
arbitrary features, for example a coordinate system which we can
choose arbitrarily, or the gauge in electrodynamics." The example
of Jacobi's principle shows that this is only the superficial
explanation for the appearance of constraints and associated
arbitrary functions.  The real reason for the Hamiltonian
constraint has nothing whatever to do with the use of a
mathematical framework containing arbitrary features.  The
underlying physical origin of the constraint is the fact that the
initial condition for a geodesic is a point in some space and a
\emph{direction} at that point.  We have made this clear in our
discussion of Jacobi's principle.  Once this has been understood,
it is evident that the corresponding Hamiltonian constraint
generates real physical change. This conclusion does not at all
mean that all constraints generate real change.  Dirac's proof is
clearly valid in many familiar cases. What this example does show
is that each constraint needs to be examined on its merits and the
underlying reason for its occurrence established.

This will be the guiding principle of our next paper, but it may
be helpful to say something here already about the relativistic
particle, which is also governed by a quadratic constraint. Given
any geodesic principle in a space of $n$ dimensions, one can
always take one of the variables (at least for a finite part of
the evolution) as the independent variable. One then obtains an
unconstrained system with $n-1$ dependent variables. Looked at
from this point of view, one may say that the Hamiltonian
constraint tells us that our system has one fewer `true degrees of
freedom' than the original $n$. This is true for both the
relativistic particle and Jacobi's principle. In this respect,
quadratic first-class constraints are no different from other
first-class constraints: they tell us that the considered system
can (generally with considerable difficulty) be described as a
system with a reduced number of degrees of freedom.

In special relativity it is also true that at any two points on a
particle worldline the particle states can be transformed into each
other exactly by means of translation and boost symmetries. There is
therefore a sense in which the state of a relativistic particle is
physically equivalent to a single `fiducial' state:
${\textbf{q}}=(0,0,0,0),~{\textbf{p}}=(1,0,0,0)$. In fact, this
emphasizes how untypical, in being too simple, the relativistic
particle is for the purposes of discussing the significance of
Hamiltonian constraints. Even though the Jacobi action has similar
symmetries, it is quite impossible to reduce the possible Jacobi
states to one in this manner.

Moreover, this does not address the issue of what the Hamiltonian
constraint \emph{does}. If we are realistic and consider a particle
in a nonempty Minkowski space, it certainly occupies observationally
different positions at different points along its worldline
irrespective of the inertial frame in which it is viewed. Further,
translation along the geodesic is driven by a Hamiltonian
constraint. Indeed, a relativistic particle interacting with an
external scalar field has essentially the same form of the action as
for Jacobi's principle. We therefore conclude that, although all
constraints indicate that we can in principle describe the system
with fewer degrees of freedom, there are constraints that generate a
physical change of the state of the system whereas other constraints
-- the ones for which Dirac's theorem holds -- do not.
%
%
\begin{acknowledgments}
We thank Hans Westman for noting that parametrized dynamics is not
different from Jacobi's principle with regard to the reality of
the change generated by its Hamiltonian constraint. We also thank
Hans, Domenico Giulini, Sean Gryb and Claus Kiefer for valuable
comments on our draft. This research was supported by a grant from
The Foundational Questions Institute (http://fqxi.org).
%
\end{acknowledgments}
%
%

%
\end{document}